\documentclass[10pt,a4paper]{article}

\usepackage{jcappub}
\usepackage{graphics}

\newcommand{\bea}{\begin{eqnarray}}
\newcommand{\eea}{\end{eqnarray}}
\newcommand{\beq}{\begin{equation}}
\newcommand{\eeq}{\end{equation}}
\newcommand{\phicmb}{\phi_\text{\tiny CMB}}
\newcommand{\Nreq}{\ensuremath{N_\mathrm{req}}}
\newcommand{\Mpl}{M_{\rm P}}
\newcommand{\kpiv}{k_{\rm piv}}

\title{Observable gravitational waves from inflation with small field
excursions}

\author[a]{Shaun Hotchkiss,}
\author[b]{Anupam Mazumdar}
\author[c]{and Seshadri Nadathur}

\affiliation[a]{Department of Physics, University of Helsinki and Helsinki
Institute of Physics, P.O. Box 64, FIN-00014 University of Helsinki, Finland}
\affiliation[b]{Physics Department, Lancaster University, Lancaster LA1 4YB,
UK\\
Niels Bohr Institute, Copenhagen, Blegdamsvej-17, Denmark}
\affiliation[c]{Rudolf Peierls Centre for Theoretical Physics, University of
Oxford, Oxford OX1 3NP, UK}

\emailAdd{shaun.hotchkiss@helsinki.fi}
\emailAdd{a.mazumdar@lancaster.ac.uk}
\emailAdd{seshadri@thphys.ox.ac.uk}

\abstract{The detection of primordial gravitational waves, or tensor
perturbations, would be regarded as compelling evidence for inflation. The
canonical measure of this is the ratio of tensor to scalar perturbations, $r$.
For single-field slow-roll models of inflation with small field excursions, the
Lyth bound dictates that if the evolution of the slow-roll parameter $\epsilon$
is monotonic, the tensor-to-scalar ratio must be below observationally
detectable levels. We describe how non-monotonic evolution of $\epsilon$ can
evade the Lyth bound and generate observationally large $r$, even with small
field excursions. This has consequences for the scalar power spectrum as it
necessarily predicts an enhancement in the spectrum at very small scales and
significant scale-dependent running at CMB scales. This effect has not been
appropriately accounted for in previous analyses. We describe a mechanism that
will generically produce the required behaviour in $\epsilon$ and give an
example of this mechanism arising in a well-motivated small-field model. This
model can produce $r\geq0.05$ while satisfying all current observational
constraints.}


\begin{document}
\maketitle
\flushbottom

\section{Introduction}
\label{section:introduction}

The positive detection of stochastic gravitational waves or tensor perturbations
of the metric, especially at very large angular scales, would be considered
strong evidence in favour of the inflationary paradigm~\cite{Grishchuk:1974ny}.
The Planck satellite will be able to detect the effects of primordial
gravitational waves on the CMB if the ratio of tensor to scalar perturbations,
$r\equiv {\cal P}_{grav}/{\cal P}_{\zeta}$, where ${\cal P}_{grav}$ and ${\cal
P}_{\zeta}$ denote the power spectra for tensor and scalar modes respectively,
is $r \gtrsim 0.05-0.1$~\cite{Planck}. It is however difficult to generate
detectably large tensor perturbations in a reasonable model of inflation which
is embedded within a fundamental physics framework, especially in supergravity
theories, where the cut-off of the theory is always assumed to be the
four-dimensional Planck scale $\Mpl\simeq 2.4\times
10^{18}$~GeV~\cite{Mazumdar:2010sa}.

The tensor-to-scalar ratio $r$ depends on only one of the slow-roll parameters,
$\epsilon$. In most models of inflation, $\epsilon$ increases monotonically,
from a value that is necessarily small during the observational window to
$\epsilon\sim 1$ at the end of inflation. Assuming such monotonic behaviour, an
upper bound may be placed on the value of $r$, known as the Lyth bound
\cite{Lyth:1996im}:
\beq
r\lesssim0.003\left(\frac{50}{N}\right)^2\left(\frac{\Delta\phi}{\Mpl}\right)^2
\,.
\eeq
Here, $\Delta\phi$ is the change in the inflaton field value in $N$ Hubble times
($e$-folds). Most scenarios require $N\sim50$ $e$-folds so that, with
$\Delta\phi<\Mpl$, $r$ cannot take a value large enough to ever be observed in
CMB experiments. This has  resulted in the widely held perspective that it is
\emph{impossible} for small-field models of inflation to generate observable
gravitational waves.

A significant tensor to scalar ratio, $r\sim 0.1$, can be obtained in
large-field models of inflation, such as ``chaotic
inflation''~\cite{Linde:1983gd}. In this class of models, slow-roll inflation
occurs when the inflaton vacuum expectation value (VEV) exceeds the Planck
scale, so that $\Delta\phi\sim5$-$10\;\Mpl$ is possible. However, even a gauge
singlet inflaton must have couplings to the Standard Model (SM) degrees of
freedom~\cite{Allahverdi:2010xz} which can generate corrections to any given
inflaton potential. An effective field theory treatment makes no sense above the
cut-off scale of quantum gravity; therefore a complete large field model of
inflation must necessarily also be fully embedded in a quantum gravity framework
such as string theory. While the quantum gravity framework allows the field to
take super-Planckian values consistently it also introduces many additional
degrees of freedom that must be kept under control. Some progress has been made
in this direction~\cite{Silverstein:2008sg}, but a full model is still a work in
progress~\cite{Burgess:2011fa}.\footnote{One problem for these models is that the inflaton does not couple 
to the Standard model quarks and leptons directly. Also, there are many more hidden degrees of freedom which can 
get excited after the end of inflation. This is a characteristic of any model where a closed string modulus is used as an inflaton~\cite{Cicoli:2010ha}.
It is also not clear whether inflaton couplings to background fluxes and matter fields would further ruin the flatness of the potential~\cite{Silverstein:2008sg}.}

Large tensor perturbations ($r\sim 0.01$-$0.1$) can also be obtained with
sub-Planckian field values in assisted inflation
scenarios~\cite{Liddle:1998jc,Kanti:1999vt,Jokinen:2004bp}, where multiple
scalar fields collectively drive inflation. Assisted inflation has found many
examples within field theory~\cite{Jokinen:2004bp}, and in string motivated
examples~\cite{Mazumdar:2001mm,Dimopoulos:2005ac}. However, these models require
as many as $10^{4}$-$10^{6}$ fields with the same mass to drive inflation
simultaneously, making them less attractive. 

These problems combined with the fact that the detection of gravitational waves
is a major observational goal means it is worth examining more closely whether
an observable tensor-to-scalar ratio can occur in any small-field models of
inflation with a single inflaton field. It would appear that the Lyth bound
negates this possibility. However, there are two ways to evade this bound. It
was argued in \cite{Hotchkiss:2008sa} that if we drop the requirement that a
single period of inflation both generates the observable perturbations and the
subsequent inflationary expansion then $N$ can be as low as $\sim8$ and a
single, small field inflation model \emph{can} generate an observably large
value of $r$. The only necessary condition for this is that a Taylor expansion
of the potential around the point of inflation is dominated by a linear term.
The unattractive feature of such a model is that some extra inflationary epoch
is necessary to generate the observed homogeneity of the universe. 

The only remaining possibility for evading the Lyth bound is to drop the
assumption of the monotonic evolution of $\epsilon$. All inflationary models
with $\phi\leq\Mpl$ and $N\gtrsim50$ that predict observable $r$ \emph{must}
violate this assumption. Some examples have recently been discussed in the
literature~\cite{BenDayan:2009kv,Shafi:2010,Rehman:2010,Okada:2011en,
Civiletti:2011}. 

In this paper we explore the consequences of the non-monotonic behaviour in
$\epsilon$ for the scalar power spectrum in such models. In section
\ref{sec:mechanism}, we explain how non-monotonicity of $\epsilon$ allows the
Lyth bound to be violated, what features are necessary in $\epsilon$ to also
match current observations, and what new observational consequences are
expected. We also argue that the required behaviour in $\epsilon$ is relatively
common when a constant vacuum energy is added to a potential that supports inflation and a
hybrid mechanism is included to bring inflation to an end

On CMB scales these models generically predict a non-negligible and
scale-dependent running of the power spectrum. The scale-dependence of the
running means it is not usually appropriate to match the power spectrum
parameters to the constraints from WMAP \cite{WMAP} only at the pivot scale. It
also provides a potentially distinct observational characteristic of small-field
inflation models in the case that gravitational waves are observed. In addition,
we note that these models generically provide a significant enhancement of power
on very small scales, which may result in generation of primordial black holes
(PBHs). These are features that are not present in other models which predict
large gravitational wave signals. 

Although our argument in section \ref{sec:mechanism} is relevant to any
small-field models that evade the Lyth bound in this manner, in section
\ref{sec:ourmodel} we use a specific model of the scalar potential to better
illustrate these features. This potential is different to the previous examples
considered in the literature. We show that it can lead to a tensor-to-scalar
ratio as large as $r\geq0.05$ with $\phi\leq\Mpl$ while also satisfying current
observational constraints. In section \ref{sec:embed} we discuss how this
potential may arise in a fundamental physics framework. Finally, we provide a
discussion and summary in section \ref{sec:discussion}.


\section{Large tensor to scalar ratio and non-monotonic evolution of
$\epsilon$}\label{sec:mechanism}

The production and evolution of tensor fluctuations during inflation is
equivalent to that for the fluctuations of a massless scalar field. The power
spectrum is therefore given by
\begin{equation}\label{eq:gravpower}
{\cal P}_{\rm grav}(k)=\left.\frac{1}{M_{\rm P}^2}\left(\frac{2 H^2}{\pi^2
}\right)
\right|_{k=aH}\,.
\end{equation}
This depends only on the scale of inflation, through the Hubble parameter $H$.
Therefore a successful detection of primordial gravitational waves from
inflation immediately tells us the inflationary energy scale. In terms of the
slow-roll parameters
\begin{equation}
\epsilon\equiv \frac{M_{\rm P}^{2}}{2}\left(\frac{V'}{V}\right)^{2}\,,~~~~~\eta
\equiv M_{\rm P}^{2}\frac{V''}{V}\,,
\end{equation}
the power spectrum and spectral index of scalar fluctuations produced during
inflation is given, to first order in the slow-roll parameters, by
\begin{equation}\label{eq:scalpower}
{\cal P}_\zeta(k)=\left.\frac{1}{M_{\rm
P}^2}\left(\frac{H^2}{8\pi^2\epsilon}\right)
\right|_{k=aH}\,\,\,\,\,\, \text{and}\,\,\,\,\,\,\,\, 
n_s=\frac{\mathrm{d}\ln{\cal P}_\zeta(k)}{\mathrm{d}\ln k}=1+2\eta-6\epsilon\,.
\end{equation}
Additionally, the running of the scalar spectral index is, to leading order in
the slow-roll parameters, given by
\beq
\alpha\equiv\frac{dn_s}{d\ln k}=-16\epsilon\eta+24\epsilon^2+2\xi^2\;,
\eeq
where $\xi\equiv\Mpl^4V^\prime V^{\prime\prime\prime}/V^2$. The ratio of tensor
to scalar fluctuations, $r$, and the spectral index of the tensor fluctuations
are given by the simple expressions
\begin{eqnarray}
r\equiv \frac{{\cal P}_{\rm grav}}
{ {\cal P}_{\zeta}}=16\epsilon~,\quad 
n_t=\frac{\mathrm{d}\ln{\cal P}_{\rm grav}(k)}{\mathrm{d}\ln k}\simeq
-2\epsilon\,.
\end{eqnarray}
These expressions depend only on $\epsilon$. Therefore, if they could both be
observed they provide a very useful consistency condition for single-field,
slow-roll inflation. It is expected that Planck could detect gravity waves if
$r\gtrsim 0.1$. An ideal future CMB polarisation experiment would be able to
detect $r\gtrsim10^{-2}$. However the tensor spectral index will be hard to measure
because of secondary anisotropies at smaller scales. 

The number of $e$-foldings of inflation, defined as $N=\int H dt$, that occur
between the field value $\phicmb$ and the end of inflation can be written as 
\begin{equation}\label{eq:efold}
 N=\int_{\phicmb}^{\phi_{\rm e}}
\frac{d\phi}{\sqrt{2\epsilon}}=\int_{\phicmb}^{\phi_{\rm e}}
\sqrt{\frac{8}{r}}d\phi\;,
\end{equation}
where $\phi_{\rm e}$ is the field value at which inflation ends and we have used the
slow-roll approximation $\dot{\phi}/H\simeq\sqrt{2\epsilon}$. If $\epsilon$ is
fixed or increasing throughout inflation this can be translated into a bound on
$r$ at horizon scales~\cite{Lyth:1996im},
\begin{equation}\label{eq:Lythbound}
r_{\text{\tiny HOR}}<0.003\left({50}/{N}\right)^2\left({\Delta \phi}/
{M_{\rm P}}\right)~.
\end{equation}
This is the Lyth bound as discussed above. It is clear that if we require
$N\geq50$ and $\Delta \phi\leq\Mpl$ then $r\leq10^{-3}$, suggesting that no single field model of inflation with small field excursion can
generate observable gravitational waves. When additional constraints provided by
considerations of naturalness and the observations of the spectral index are
included, this bound becomes even stronger \cite{Hotchkiss:2008sa}. 

Note that the current observational window corresponds to only $\sim 8$ of the
total number of $e$-folds we require. Using $N=8$ in eq.~\eqref{eq:Lythbound} gives the
bound $r_{\text{\tiny HOR}}\lesssim0.1$. Thus if the other $\sim 40$ $e$-folds
of inflation can be generated by some other mechanism, then single-field
inflation with small field excursions can generate an observable tensor signal
without violating the Lyth bound~\cite{Hotchkiss:2008sa}. On the other hand, the
smallness of the observational window also allows us to exploit another loophole
in the derivation of the Lyth bound, namely the assumption that $\epsilon$
increases monotonically. This seems a natural assumption given that during
inflation $\epsilon \ll 1$ and at the end of inflation $\epsilon \sim 1$;
current observational constraints also strongly favour $\epsilon$ increasing
during the $\sim8$ $e$-folds of the CMB observational window. However, outside
this window the behaviour of $\epsilon$ is not constrained and the assumption of
monotonicity is not strictly necessary. Relaxing this assumption makes it
possible to construct a scalar potential for a single field that violates the
Lyth bound.

This loophole was first exploited in \cite{BenDayan:2009kv}, where a fifth-order
polynomial form for the scalar potential was used to generate observably large
$r$ (as large as $r=0.1$) with $\sim 60$ $e$-folds of inflation, while matching
the WMAP constraints on the amplitude, spectral tilt and running of the scalar
power spectrum \cite{WMAP} \emph{at the pivot scale}, all with field excursion
$\Delta \phi\lesssim \Mpl$. Supersymmetric models using a similar principle but with an
additional hybrid mechanism to bring inflation to an end have also been
studied~\cite{Shafi:2010,Rehman:2010,Okada:2011en,Civiletti:2011}, where the
relevant contributions to the potential arise from quadratic and quartic terms
in the inflaton field $\phi$ with opposite coefficients. These were matched to
observational constraints from WMAP in a similar manner, and values of $r\sim
0.03$ were obtained.

However, the behaviour of $\epsilon(\phi)$ that is necessary in order to evade
the Lyth bound and satisfy observational constraints in a complete inflationary
model can be quite complicated, as we discuss below. In general it is not
sufficient to simply match the parameters of the scalar power spectrum at the
WMAP pivot scale, as done in
\cite{BenDayan:2009kv,Shafi:2010,Rehman:2010,Okada:2011en,Civiletti:2011}.


\subsection{The necessary features in $\epsilon$}\label{sec:epsfeat}

To violate the bound in eq.~\eqref{eq:Lythbound} $\epsilon$ must decrease at
some point during the inflationary epoch.\footnote{This is clear from
eq.~\eqref{eq:efold} where, if $\epsilon$ decreases, $\Delta N$ will increase
for the same field excursion, $\Delta \phi$.} Unfortunately this alone is not
enough to successfully evade the Lyth bound and match all observations. To
achieve both, a complete small-field model must have the following features:
\begin{enumerate}
 \item At CMB scales $\epsilon$ must be large enough to generate an observable
value of $r$.
 \item $\epsilon$ must \emph{increase} over the $\sim8$ $e$-fold observational
window.
 \item After observable scales have left the horizon, $\epsilon$ must quickly
decrease.
 \item $\epsilon$ must eventually increase again to end inflation.
\end{enumerate}
Condition (2) is dictated by a combination of the spectral index constraint from
WMAP~\cite{WMAP} and the observed value of $\sigma_8$ from large-scale structure
(e.g. galaxy clusters, see~\cite{Mantz:2009fw}), which means the spectrum must
decrease over the observational window.\footnote{When running of the spectral
index is allowed, the best fit value for the spectral index is indeed $n_s>1$,
but the running is $\alpha<0$, therefore $\epsilon$ must still increase
eventually. Also, the value of $\sigma_8$ measured independently from LSS
strongly favours a primordial spectrum that decreases in amplitude between WMAP
and LSS scales.} The quick decrease of $\epsilon$ required in condition (3) is
necessary to generate enough $e$-folds of inflation. If instead $\epsilon$
decreases gradually, it will need to eventually decrease to a much smaller
value because $\epsilon\propto\Delta \phi/\Delta N$
and we require $\Delta \phi\leq\Mpl$. The change in $\epsilon$ must also not be too sharp
in order to not violate slow-roll. Note that a \emph{complete} model of
inflation also requires a stable vacuum state in which $V'\rightarrow 0$.

Achieving all of these conditions with a single potential is difficult, and a
viable model would require a fifth-order polynomial form for the potential, as
in \cite{BenDayan:2009kv}. The most demanding aspect of the conditions is that
the single scalar potential \emph{both} dramatically reduces $\epsilon$ after
observable scales leave the horizon and subsequently dramatically increases
$\epsilon$ to end inflation. If inflation is instead brought to an end by
another mechanism, such as a hybrid transition, as in
\cite{Shafi:2010,Rehman:2010,Okada:2011en,Civiletti:2011}, condition (4) is not
required. This makes the task of constructing the potential simpler.

In fact, the condition that $\epsilon$ first increases and then decreases is not
uncommon and can be achieved by simply adding a constant vacuum energy term to 
a potential that already supports inflation (and has increasing $\epsilon$). Before
the constant term becomes completely dominant $\epsilon$ will continue to increase as before. 
Then, when the constant term does come to completely dominate,
$\epsilon\rightarrow 0$. For a large enough vacuum term, this transition will occur before $\epsilon >1$. Of course to successfully evade 
the Lyth bound as described above, the occurrence of this feature must coincide with the end of 
the observational window and occur with the right magnitude, which is not guaranteed. However, the non-monotonic behaviour of 
$\epsilon$ will be. Refs.~\cite{Shafi:2010,Rehman:2010,Okada:2011en,Civiletti:2011} are examples of
this mechanism in action, as is our model in section \ref{sec:embed}.


\subsection{Typical observational features}\label{sec:constraints}

Unlike most models of inflation, models with non-monotonic evolution of
$\epsilon$ will not reproduce an exact power-law spectrum of primordial density
perturbations. Apart from an observable tensor-to-scalar ratio, which is
obtained by design, such models generically exhibit the following two features:
\begin{enumerate}
 \item A non-negligible, scale-dependent running, of the scalar spectral index.
 \item A significant increase in the primordial power spectrum on very small
scales.
\end{enumerate}
  
The non-negligible running arises because there must be significant evolution of
$\epsilon$ while observable scales are crossing the horizon. While it is
possible in principle for all the evolution of $\epsilon$ to occur only after
the observable scales have crossed the horizon, this requires \emph{very} sharp
features in $\epsilon$ since, as mentioned above, the later the decrease in
$\epsilon$ begins, the greater the over-all decrease must be in order to
maintain inflation for sufficient $e$-folds. If $\epsilon$ decreases too much,
the perturbative treatment of the scalar fluctuations, and hence the consistency
of inflation, breaks down.\footnote{This can only be overcome by a concurrent
sharp drop in the energy scale of inflation. Such a sharp change would be more
suitably described as an inflationary model with multiple periods of inflation
\cite{Adams:1997de}, rather than one period with an evolving $\epsilon$.} Even
if the parameters of the model are chosen in order to impose a small running
\emph{at the WMAP pivot scale} (as in
\cite{BenDayan:2009kv,Shafi:2010,Rehman:2010,Okada:2011en,Civiletti:2011}) the
necessary evolution of $\epsilon$ required for consistency means that the
running cannot remain small over the observational window, i.e., the running
will be \emph{scale-dependent}. 

This means the description of the scalar power spectrum using the simple
power-law model is not appropriate for models implementing such an evolution of
$\epsilon$ to evade the Lyth bound. As a result, the value of $\sigma_8$
actually measured from LSS observations would be different to that inferred from
the CMB if the data were analysed assuming a power-law spectrum. 

Finally, the increase in the spectrum at very small scales is a result of the
necessary decrease of $\epsilon$ outside the observational window. If the
amplitude of the spectrum on these scales is large enough, primordial black
holes (PBHs) will be formed. Current cosmological constraints on PBH's restrict
$\mathcal{P}_\zeta\lesssim 10^{-2}$ (see e.g.
\cite{Carr:2009jm,Alabidi:2009bk,Drees:2011hb}).

Currently, none of these features have definitely been observed, which serves to
place constraints on what small-field models can circumvent the Lyth bound. On
the other hand, if primordial gravitational waves were to be observed,
subsequent detection of any of these features would allow us to distinguish
between small-field models and other inflationary models which can generate
large tensor signals.


\section{A well-motivated potential}\label{sec:ourmodel}

In the previous section we have discussed in very general terms the conditions
that must be satisfied by any small-field model of inflation that circumvents
the Lyth bound to produce a large tensor signal. We now illustrate these
properties with a concrete example of a well-motivated scalar potential that can
be obtained from a fundamental physics framework. 

A model of inflation based on gauge invariant MSSM (minimal supersymmetric Standard Model) flat-direction fields was
introduced in \cite{Allahverdi:2006iq}, where inflation occurs about a point of
inflection in the potential. In all of these models the inflaton carries the Standard Model charges and inflation ends in a vacuum where 
the MSSM gauge group is restored. Therefore the inflaton decay naturally excites all the MSSM degrees of freedom~\cite{Allahverdi:2011aj}.

This model was subsequently extended to include a constant vacuum energy term arising from the next-to-minimal supersymmetric Standard Model (NMSSM)
\cite{Hotchkiss:2011am,Enqvist:2010vd},\footnote{The constant term in this potential could also arise from the vacuum energy density present within a gravity mediated supersymmetry 
breaking sector~\cite{Mazumdar:2011ih}, or within the MSSM-landscape~\cite{Mazumdar:2011xe}. These require a mechanism other than a hybrid transition to end inflation and remove the vacuum term.} such that the potential
was stable under radiative corrections and did not require excessive fine-tuning
to reproduce a power spectrum consistent with observations. 

The potential in this model can be easily parameterized by
\begin{equation}\label{pot}
 V(\phi)=V_0 +A\phi^p-B\phi^q+C\phi^s\;,
\end{equation}
while the hybrid field remains trapped in the false vacuum. Although in
principle the three powers in the potential can take many values, for the
purpose of illustration here we will use $p=2$, $q=6$ and $s=10$. We will assume
here that the hybrid transition brings inflation to an end after the appropriate
number of $e$-folds, so that no constraint is required on the value of
$\epsilon$ at the end of inflation. The details of the embedding of this
potential into a model of particle physics imposes constraints upon the the
values of $V_0$, $A$, $B$ and $C$. In this section we first deal with the model
as a general framework, and then point out one potential embedding of the model
in section \ref{sec:embed}. 

To start with, the observables we choose our model to reproduce are: the
amplitude of the scalar power spectrum, $A_k$, the scalar spectral
index, $n_s$ and the tensor to scalar ratio, $r$, at the WMAP pivot scale
$\kpiv=0.002\;\rm{Mpc}^{-1}$. In section \ref{sec:beyondSR} we examine the
spectrum produced by this model beyond this power-law assumption. For a given
potential $V(\phi)$, when the field is slowly rolling and takes field value
$\phicmb$ these observables are given by
\begin{equation}
 A_k=\frac{V(\phicmb)}{24 \pi^2 \epsilon(\phicmb)},
\end{equation}
\begin{equation}
 n_s=1+2\eta(\phicmb)-6\epsilon(\phicmb)
\end{equation}
and
\begin{equation}
 r=16\epsilon(\phicmb).
\end{equation}
We choose $\phicmb$ to be the field value at which the pivot scale $\kpiv$
crosses the horizon. Conveniently, these equations can be re-arranged to give
$V(\phicmb)$, $V'(\phicmb)$ and $V''(\phicmb)$ as a function of the observable
quantities
\begin{equation}
 V(\phicmb)=\frac{3}{2}(A_k r\pi^2),
\end{equation}
\begin{equation}
 V'(\phicmb)=\frac{3}{2}(A_k r\pi^2)\sqrt{\frac{r}{8}}
\end{equation}
and
\begin{equation}
 V''(\phicmb)=\frac{3}{4}\left(\frac{3r}{8}+n_s-1\right)(A_k r \pi^2).
\end{equation}
Therefore, for any chosen $\phicmb$ and $V_0$, we obtain a matrix equation for
the variables $A$, $B$ and $C$:
 \begin{equation}\left(\begin{tabular}{ccc}
 $\phicmb^2$ & $-\phicmb^3$ & $\phicmb^4$\\
$2\phicmb$ & $-3\phicmb^2$ & $4\phicmb^3$\\
$2$ & $-6\phicmb$ & $12\phicmb^2$
      \end{tabular}\right)
\left(\begin{tabular}{c}
      $A$\\
      $B$\\
      $C$\\
      \end{tabular}\right)
\begin{tabular}{c}
      \\
      =\\
      \\
      \end{tabular}
\left(\begin{tabular}{c}
      $V(\phicmb)-V_0$\\
      $V'(\phicmb)$\\
      $V''(\phicmb)$\\
      \end{tabular}\right).
\end{equation}
This is easily inverted to find the required values of $A$, $B$ and $C$, and
thus the complete potential, $V(\phi)$. Note that the slow-roll constraints
$\epsilon\ll1$ and $|\eta|\ll1$ are automatically satisfied at $\phicmb$, but
there is no guarantee that the generated potential will be able to generate the
minimum required number of $e$-folds of inflation, \Nreq. This must be checked
for each chosen value of $\phicmb$ and $V_0$. There is also no guarantee that
for all \Nreq~$e$-folds $A_k$, $n_s$ and $r$ will remain close to their values
at $\phicmb$.

\begin{figure}[t]
\center
 \includegraphics[scale=0.3]{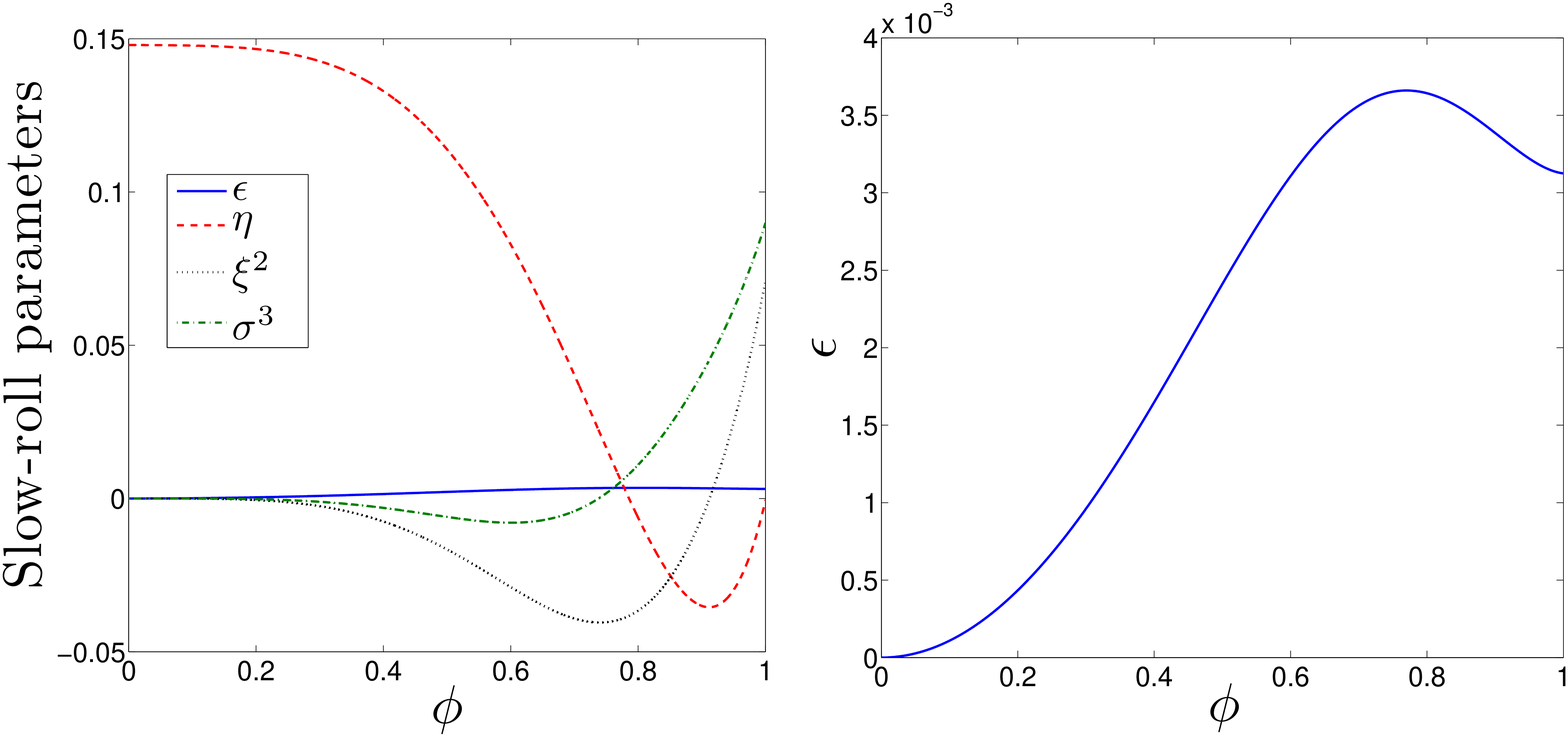}
 \caption{\label{fig:slowroll} \emph{Left panel}: The behaviour of the slow-roll
parameters as a function of inflaton field value $\phi$, measured in units of
$\Mpl$, for the model parameters given in equation~(\ref{sample}). \emph{Right
panel}: $\epsilon$ as a function of $\phi$ for the same parameters, shown with a
magnified scale.}
\end{figure}

Using this method, we have reduced the numerical difficulty to a two-dimensional
problem. Having chosen the observational parameters we wish to match, we need
only scan over the two parameters $V_0$ and $\phicmb$ to find a potential that
can sustain inflation for $>\Nreq$~$e$-folds. The waterfall transition that ends inflation is then restricted to occur exactly $\Nreq$~$e$-folds after $\phi=\phicmb$. As a test of the method we will show that it is possible to obtain:
\begin{equation}
r=0.05,~~n_s=0.98,~~A_k(\kpiv)=2.3\times 10^{-9},~~{\rm for}~\Nreq=50,~~{\rm
and}~~\phicmb=\Mpl. 
 \end{equation}
The values of $n_s$ and $A_k$ are the WMAP 7-year best fit values for a
power-law scalar spectrum with non-zero tensor spectrum \cite{WMAP}. As noted
above, the power spectrum thus produced will generically differ significantly
from a power-law form. Therefore we further require that the power at very small
scales is always $\mathcal{P}(k)\leq10^{-2}$ during inflation, to ensure there
is no over production of PBH's~\cite{Carr:2009jm}. The final constraint imposed
is that the power at $k=0.3$ Mpc$^{-1}$ is not more than $10\%$ different than
that for a power-law spectrum with $n_s=0.98$. Note that $k=0.3$ Mpc$^{-1}$
corresponds to multipole $l>3000$. Although the precise scale at which
this matching is imposed is somewhat arbitrary, this constraint is necessary to
ensure that the model does not predict too much power on intermediate scales,
which would conflict with LSS observations of $\sigma_8$~\cite{Mantz:2009fw} and
measurements of the small scale angular power spectrum of the CMB from, e.g.,
QUaD \cite{Gupta:2009sy}, ACBAR \cite{Reichardt:2008ay} and Boomerang
\cite{MacTavish:2005yk}. 
 
A particular choice of $V_0$ which satisfies all of these constraints produces
the following parameters:
\begin{equation}
\label{sample}
V_0^{1/4}=0.0063\Mpl,~~A=1.19\times10^{-10}\Mpl^2,~~ B=2.87\times
10^{-11}\Mpl^{-2}\,,~~ C=6.90\times10^{-12}\Mpl^{-6}. 
\end{equation}
For these parameters, figure~\ref{fig:slowroll} shows the behaviour of the
slow-roll parameters $\epsilon$, $\eta$, $\xi^2$ and
$\sigma^3\equiv\Mpl^6V^{\prime2}(d^4V/d\phi^4)/V^3$ as a function of $\phi$. It
is clear that the required evolution of $\epsilon(\phi)$, outlined in section
\ref{sec:epsfeat}, is satisfied. Note that the ordinary slow-roll hierarchy is
not maintained due to the non-monotonic behaviour of $\epsilon$, and that the
other slow-roll parameters themselves show non-monotonic behaviour. The power
spectrum will therefore differ from the standard power-law form, and in general
the slow-roll approximation for calculating the power spectrum will not be
valid.

\subsection{Predictions beyond slow-roll and a power-law spectrum}
\label{sec:beyondSR}

\begin{figure}[t]
\center
 \includegraphics[scale=0.3]{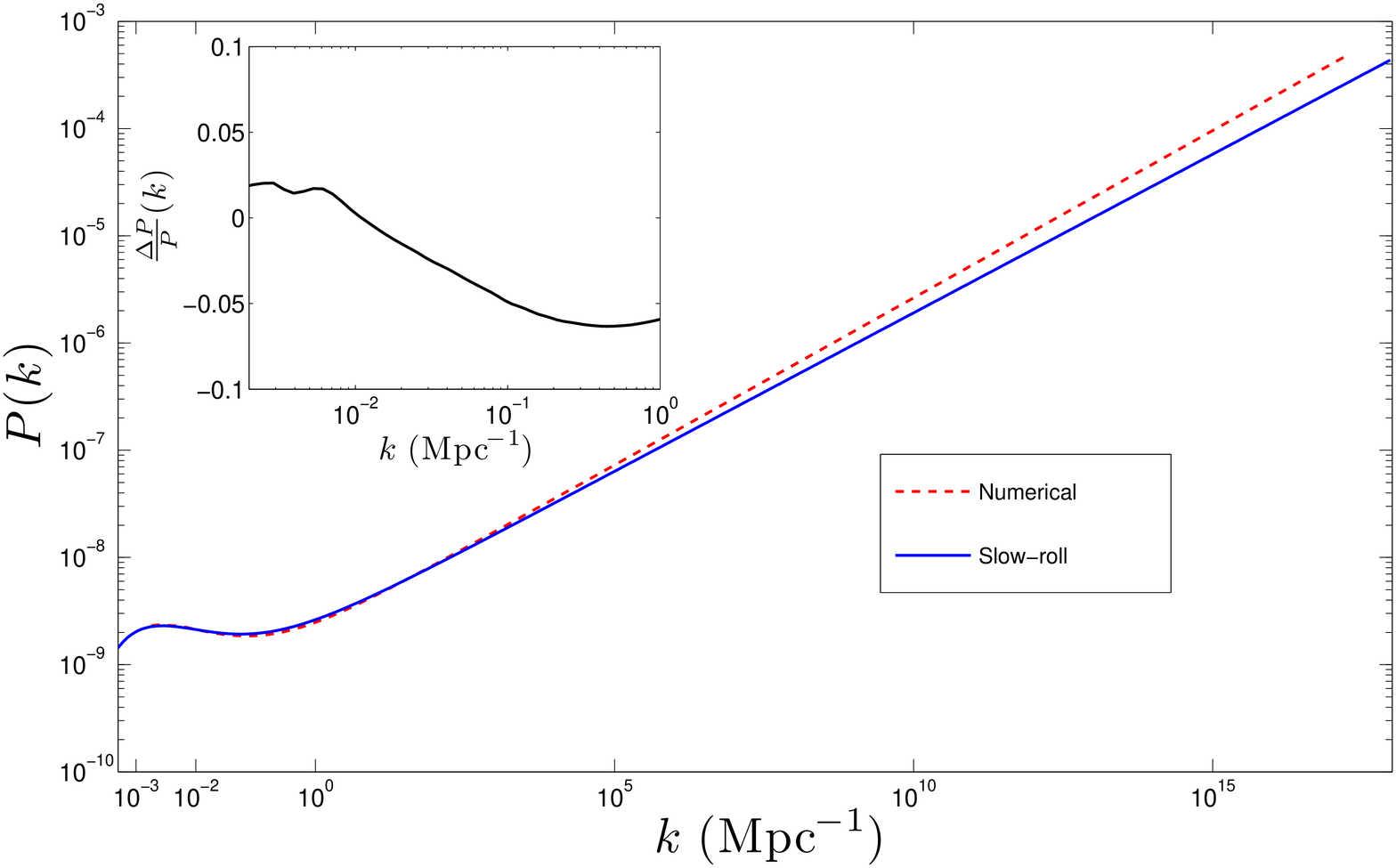}
\caption{\label{fig:power} Power $P(k)$ for the model parameters of
equation~\eqref{sample} for the entire range of scales $k$ that cross the
horizon during the \Nreq~$e$-folds of inflation. The solid blue curve is
obtained from the slow-roll approximation and the red dashed curve from a full
numerical calculation. \emph{Inset}: The fractional difference
$(P_\mathrm{num}-P_\mathrm{s.r.})/P_\mathrm{num}$ between the numerical
calculation and the slow-roll approximation at large scales corresponding to the
current observational window.}
\end{figure}

The predicted power spectrum obtained from the slow-roll approximation will be
correct to the same order as the slow-roll parameters themselves. Therefore, if
the higher order slow-roll parameters $\xi^2, \sigma^3$ etc. are generically
large, the corrections to the predicted power spectrum might also be large. In
order to check this we calculate the spectrum numerically. To do this we assume
the Bunch-Davies vacuum state and numerically integrate the scale factor, field
value and curvature perturbation from deep inside the horizon until well after
horizon exit using the equation of motion given by linear perturbation
theory~\cite{Mukhanov:1990me}. Figure \ref{fig:power} shows $\mathcal{P}(k)$
calculated by the two different methods. At small scales the deviation is
significant. Even at large scales, as shown in the inset to
figure~\ref{fig:power}, these differences are of order $5\%$ or more, and for
different choice of model parameters may be more significant than this. For all
subsequent plots we therefore use the full numerical calculation of
$\mathcal{P}(k)$ rather than the slow-roll approximation.

\begin{figure}[t]
\center
 \includegraphics[scale=0.3]{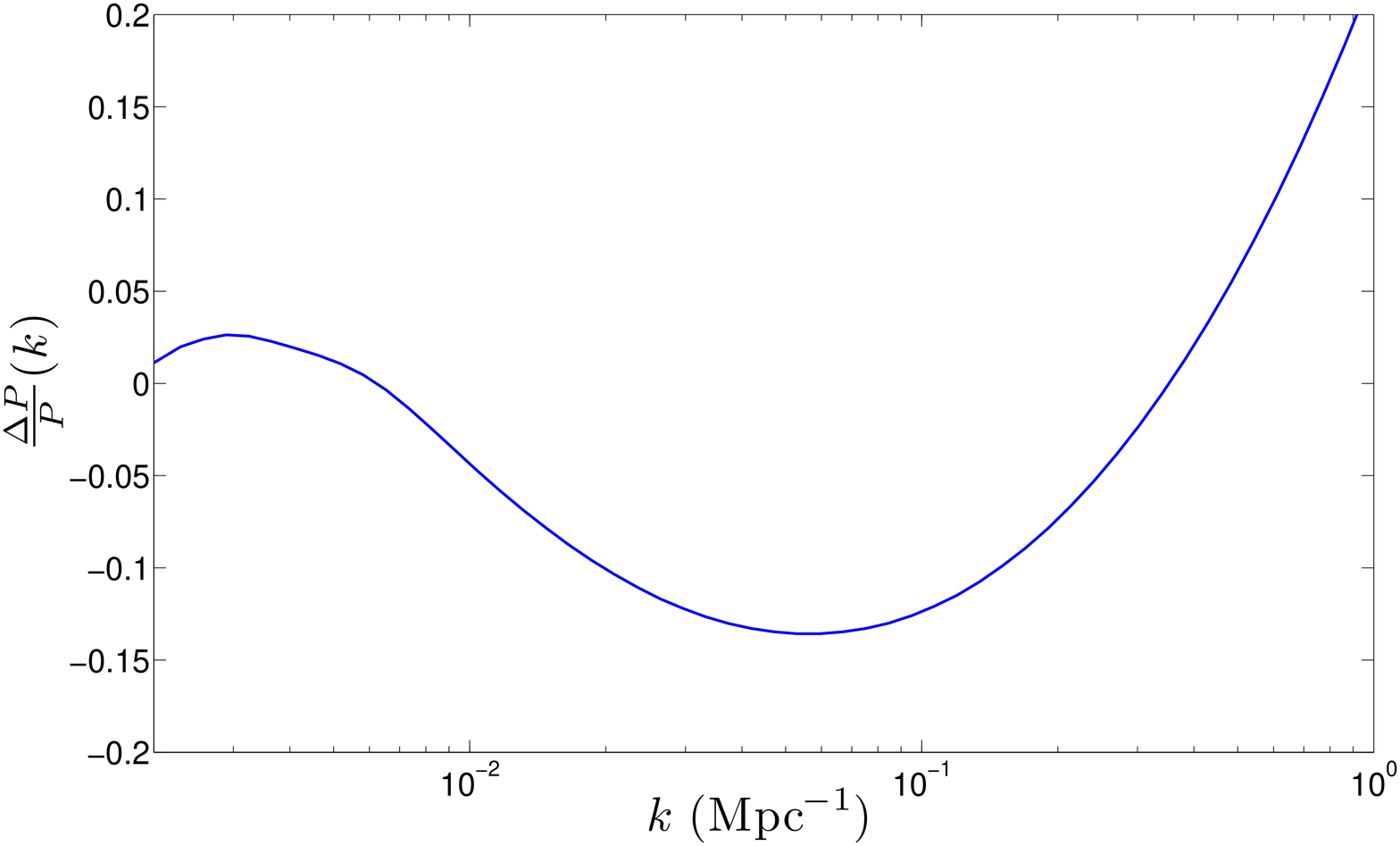}
\caption{\label{fig:diffpower} The fractional difference between the numerically
evaluated power spectrum $P(k)$ for parameters in equation~\eqref{sample} and
the assumed power-law form with the values of $A_k=2.3\times10^{-9}$ and
$n_s=0.98$ at the pivot scale $\kpiv=0.002$ Mpc$^{-1}$ to which the model was
matched as described in the text.}
\end{figure}


It is also instructive to compare the power spectrum obtained from the full
numerical calculation for this model with the power-law spectrum with
$A_k=2.3\times10^{-9}$ and $n_s=0.98$. These were the parameters of the power
spectrum to which we had matched our model at the WMAP pivot scale. Despite this
matching at $\kpiv$ and the additional matching (to within $10\%$) at $k=0.3$
Mpc$^{-1}$, the fractional difference in the power spectrum compared to the
assumed power-law form can become significant over the range of scales
constrained by current CMB measurements, as shown in figure~\ref{fig:diffpower}.
This shows that a power-law is not the ideal framework with which to analyse the
power spectrum generated in this model.

For the parameters we have chosen in eq.~\eqref{sample}, the two spectra match
to within $20\%$ over the WMAP constrained range. This is equivalent to the
accuracy quoted in the parameter $A_k$ when both tensors and running are
included in the WMAP analysis \cite{WMAP}. Therefore such a form of the
primordial power will be within current constraints, but should be possible to
distinguish when using the tighter constraints expected from Planck.

\begin{figure}[t]
\center
 \includegraphics[scale=0.3]{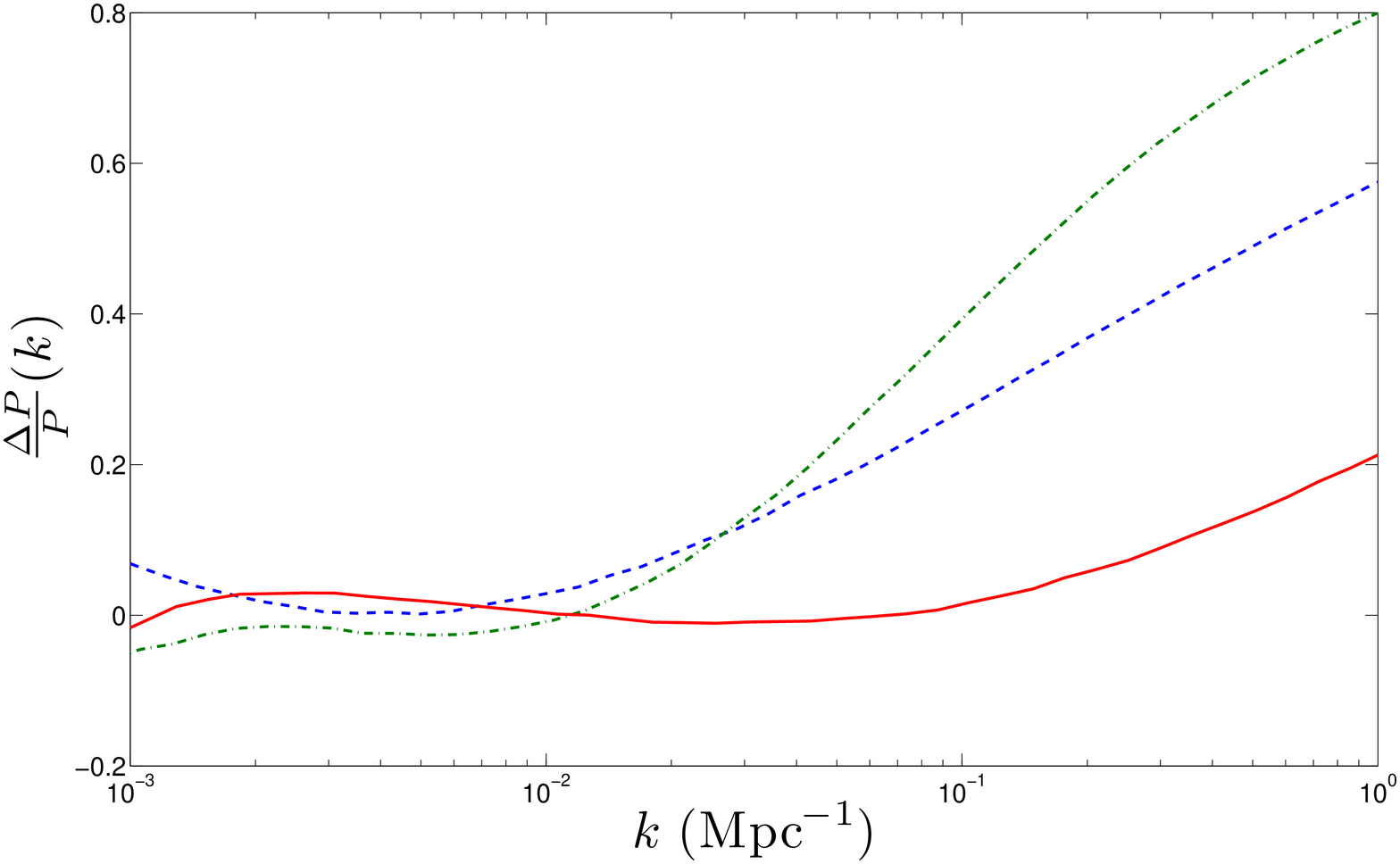}
\caption{\label{fig:comparison} The fractional difference between the
numerically evaluated power spectrum $P(k)$ and the assumed power-law form to
which the values of $A_k$, $n_s$ and $\alpha$ are matched at the pivot scale
$\kpiv=0.002$ Mpc$^{-1}$. The green dash-dot curve is for model 1, the blue
dashed curve for model 2 and the red solid curve for model 3. The various models
are described in section \ref{sec:beyondSR}. Model 1 is taken from
\cite{BenDayan:2009kv} and model 2 from \cite{Rehman:2010}. Note that although
the power spectra are matched to the power-law form at the pivot scale, they all
necessarily deviate from this form at other scales within the observational
window. Only model 3 matches the power-law to within current observational
accuracy.}
\end{figure}

This deviation from a power-law form of the power spectrum is an important
feature of our model and in fact is a generic feature of \emph{any} small-field
model generating observable gravitational waves. This has not been accounted for
in previous analyses of small-field models of inflation generating large $r$. To
demonstrate this, we consider two different models from the literature. The
first model is based on the fifth-order polynomial form for $V(\phi)$ considered
in \cite{BenDayan:2009kv}, which gives a tensor-to-scalar ratio $r=0.02$. The
parameter values are taken from Table 1 in \cite{BenDayan:2009kv} (second line)
and have been chosen in order to match the power spectrum to that of a power law
with $A_k=2.3\times10^{-9}$, $n_s=0.99$ and $\alpha=0.001$ at the pivot scale
$\kpiv$. We refer to this as model 1. The second model, referred to as model 2,
is based on the model of supersymmetric hybrid inflation considered in
\cite{Rehman:2010}. The relevant portion of the potential for this model is 
\beq
\label{shafipot}
V(\phi)=V_0\left(1+\frac{|\kappa_s|}{\Mpl^2}\phi^2+\frac{\gamma_s}{2\Mpl^4}
\phi^4\right)
\eeq
at the time of generation of the CMB perturbations (for more details, see
\cite{Rehman:2010}). We choose the parameter values for this potential in order
to obtain $r=0.02$, $A_k=2.3\times10^{-9}$, $n_s=0.96$ and $\alpha=0.003$ at
$\kpiv$ (which is assumed to cross the horizon at $\phicmb=0.5\Mpl$). This
results in $V_0=6.7\times10^{-10}\;\Mpl^4$, $|\kappa_s|=0.08$,
$\gamma_s=-0.118$. Note that the parameters of these models are chosen as
examples which both generate the same value of $r$. In
figure~\ref{fig:comparison}, we plot as a function of the scale $k$ the
fractional difference between the numerically evaluated power spectrum
$\mathcal{P}(k)$ for each of these models and the power-law form of the power
spectrum (with scale-independent running) to which they are matched at the pivot
scale. For comparison, we include the same plot for parameter values
\begin{equation}
\label{sample2}
V_0^{1/4}=0.0051\Mpl,~~A=3.54\times10^{-11}\Mpl^2,~~ B=1.75\times
10^{-11}\Mpl^{-2}\,,~~ C=8.81\times10^{-12}\Mpl^{-6}. 
\end{equation}
for our model of equation \eqref{pot} (here $\phicmb=0.8\;\Mpl$). These
parameter values are chosen to obtain $r=0.02$, $A_k=2.3\times10^{-9}$,
$n_s=0.96$ and $\alpha=0.001$ at $\kpiv$ and thus correspond closely to model 2.
We refer to this as model 3.


From figure \ref{fig:comparison} it is obvious that all three models result in
scalar power spectra that deviate from the power-law form at intermediate or
small scales even when they are explicitly matched to a power-law spectrum at
large scales. As explained in section~\ref{sec:constraints}, this is to be
expected as a generic feature of any small-field model generating observable
gravitational waves. It is therefore not appropriate to merely match the power
spectrum to a power-law at $\kpiv$: both models 1 and 2, which have been claimed
to match observational constraints on the scalar power spectrum, in fact predict
a significant excess of power at scales of $k\simeq0.1$ Mpc$^{-1}$, which will
put them in tension with $\sigma_8$ measurements. Note that model 3, although
certainly not a power-law, is the only model of the three that matches the
power-law to within current observational accuracy.

If a large $r$ is observed along with a large running, then it will be possible
to directly compare the goodness of fit of our model's spectrum and a simple
running power-law. For the running power-law there are four free parameters.
These are $r$, $A_k$, $n_s$ and $\alpha={\rm d} n_s/{\rm d}\ln k$. For the model
presented in this paper there also four parameters. These are $V_0$, $A$, $B$
and $C$. In principle, because our model requires other physics to end inflation
there is one additional parameter. This is either $\phicmb$ or $\phi_{\rm e}$.
However, the extra constraints we impose that $\phicmb\lesssim \Mpl$ and that
the spectrum at the smallest scales is less than $10^{-2}$, strongly constrains
this extra parameter. Also, if the effects of PBH's were to be observed, this
would provide a fifth observable for our model. 

It is also interesting to note that if PBH's are generated at the correct
scales, then they can be a viable dark matter
candidate~\cite{Carr:2009jm,Drees:2011hb}. In ordinary inflation models it is
difficult to generate enough power to create PBH's, whereas in our model it is
difficult not to. Nevertheless, a full exploration of the implications of this
although interesting is well beyond the scope of the present work.


\subsection{The model embedded within supergravity using MSSM flat directions}
\label{sec:embed}

In this section we describe a simple particle theory embedding for the scalar
potential of the form of equation~(\ref{pot}). This potential can be obtained
within the MSSM where there are $D$-flat directions which are lifted by
non-renormalizable operators  (for a review see~\cite{Enqvist:2003gh}).
At high scales, the supergravity corrections dominate the soft-supersymmetry breaking contribution of the inflaton potential, 
as suggested in \cite{Mazumdar:2011ih}. The MSSM potential obtains 
leading corrections which are dominated by the Hubble-induced terms. The total potential 
is given by (see~\cite{Mazumdar:2011ih,Kasuya:2006wf}):
\begin{equation}
V= V_{0}+c_H H^{2}{\left\vert \phi \right\vert}^2 -
a_{H}\lambda_{n}H\frac{\phi^6}{\Mpl^{3}}
+ \lambda_{n}^{2}\frac{|\phi|^{10}}{\Mpl^{6}}
\end{equation}
where $V_{0}=3H^{2}\Mpl^{2}$,~$c_{H}$ and $a_{H}$ are numerical factors arising from 
supergravity corrections. We will
work with the assumption that $c_{H}, ~a_{H}\sim {\cal O}(1-100)$. Note that 
the Hubble-induced corrections match the expressions given in eqs.(\ref{pot}) and (\ref{sample}) with
appropriate values of $V_{0},~A,~B,~C$. We can ignore the soft terms such as $m_{\phi}$ as
compared to the Hubble-induced terms.\footnote{Note that the effective mass term is now governed
by the large Hubble-induced mass term. Therefore in this model there is no supergravity-$\eta$ problem.}
By comparing to our sample point, we can read off the parameters that correspond to \eqref{sample}. These are:
\begin{equation}\label{match}
H=3.5\times 10^{11}~{\rm GeV}\,,~~c_{H}\sim 8.2\,,~~a_{H}\sim
76\,,~~\lambda_{n}\sim 2.6\times 10^{-6}\,.
\end{equation}
Our analysis suggests that in order to realize a large tensor-to-scalar ratio,
$r\sim 0.05$, we would require a small value for of $\lambda_{n}$. This is the coefficient of the non-renormalizable operator
for the MSSM flat directions, such as ${\bf udd}$ and ${\bf LLe}$~\cite{Allahverdi:2006iq} (${\bf u}$ and ${\bf d}$ are the right handed squarks,  ${\bf L}$ is the left handed slepton 
while ${\bf e}$ corresponds to the right handed selectron).
The other coefficients, $c_{H}$ and $a_{H}$ are within the allowed range of supergravity corrections arising from the mixing of the K\"ahler potential and superpotential. However,
the non-renormalizable interaction, $\lambda_{n}$ turns out to be unnaturally small.
Both ${\bf udd}$ and ${\bf LLe}$ carry global $U(1)$ numbers, and in both the cases it
is expected that $U(1)$ would be broken by quantum gravity
effects with order unity, i.e. $\lambda_{n}\sim {\cal O}(1)$. However, it is
not unforeseeable that due to some reasons quantum gravity effects break
the global baryon or lepton number operators {\it softly}. Inspite of this
challenge it is still nice to see that it is possible to obtain a large tensor-to-scalar
ratio in a model where $\Delta \phi \leq\Mpl$ and the end of inflation creates
{\it solely} the MSSM degrees of freedom, therefore creating a thermal bath with
all the Standard Model quarks and leptons, and also the dark matter component as
the lightest supersymmetric particle~\cite{Allahverdi:2007vy}.

\section{Summary and Discussion}
\label{sec:discussion}

In this paper we have discussed small-field models of inflation that can
generate observably large gravitational waves. As has been noted in
\cite{Hotchkiss:2008sa,BenDayan:2009kv,Shafi:2010,Rehman:2010,Okada:2011en,
Civiletti:2011}, if a significant tensor-to-scalar ratio is observed through the
CMB this does not rule out small-field inflation. This result does not
contradict the well known Lyth bound because it does not satisfy one of the
assumptions used to derive this bound, namely that $\epsilon$ increases
monotonically during inflation. 

We have discussed the generic features that $\epsilon(\phi)$ must possess in
such models in order to meet observational constraints, and the distinct
observational signatures any non-monotonic $\epsilon$ model will produce. The
most important of these are a scale-dependent running of the scalar spectral
index and an enhancement of power on very small scales, potentially leading to
the formation of primordial black holes. As a result of the scale-dependent
running of the spectral index, the primordial power spectrum generally differs
from the power-law form usually assumed so it is not appropriate to fit the
spectrum to the data at one scale assuming such a form. Some previous examples
discussed in the literature
\cite{BenDayan:2009kv,Shafi:2010,Rehman:2010,Okada:2011en,Civiletti:2011} have
not taken this into account and so may be more constrained by CMB observations
than was previously realised.

A deviation of the primordial power spectrum from a power-law form may be
inferred from the data if $\sigma_8$ measurements from large-scale structure
were to be found to disagree with those predicted from CMB data when analysed
assuming a power-law spectrum. If indeed a tensor signal were to be observed in
the future, such a scale-dependent running and primordial black holes would be a
smoking gun signal with which to distinguish these small-field inflation models
from other models which can also generate large $r$.

We have argued that the required behaviour in $\epsilon$ is relatively common when a
constant vacuum energy is added to a potential that supports inflation and a hybrid
mechanism is included to bring inflation to an end. We have also discussed a
small-field model which implements this mechanism. We have shown the existence 
of a point in the parameter space of
this model with $r=0.05$ that predicts a scalar power spectrum $\mathcal{P}(k)$
which deviates from the current best-fit power-law by less than $20\%$ over all
the currently observed range of $k$ values.  When a tensor signal and running of
the scalar spectral index is included, this is the accuracy to which the WMAP
data can constrain the amplitude of the running power-law spectrum. We conclude
that our model will not be able to give a fit to current data that is
statistically better or worse with any significance compared to a running
power-law model, because there is not enough information in the data. However,
if a future CMB observation, perhaps from Planck, were to measure both $r$ and
$\alpha$ with statistical significance, a full Markov Chain Monte Carlo (MCMC)
analysis of our model could provide useful information.

Additionally, we observed that the generation of primordial black holes after
inflation is a generic feature of non-monotonic $\epsilon$ models. The fact that
these have not yet been observed provides tight constraints on the allowed
parameter space and effectively fully constrains one free parameter of the
model. We have not explored this observational feature beyond using it as a
constraint; however it would appear to be an interesting avenue for future
pursuits.

Although the scalar potential we have discussed satisfies current observational
constraints, embedding it in a fundamental physics model is not without
difficulty. We have provided an example of how this may occur within an MSSM
framework, but in this implementation the coefficient of one term in the
potential must take small values. The initial conditions of this
model also need to be delicately arranged if slow-roll inflation is to occur, 
possibly through the dynamical mechanism suggested in~\cite{Allahverdi:2008bt}.
This is also true of other small-field models in the literature.
Nevertheless, the fine-tuning problems of these small-field models appear to be
no greater than those arising in reconciling large-field models with fundamental
theory. Therefore, although from a theoretical perspective large $r$ values may
be disfavoured, if a primordial tensor signal were in fact to be observed then
contrary to popular wisdom small-field models should be considered at least as
seriously as large-field models. Thankfully, due to their distinct observational
signatures, the choice between them then need not be based on theoretical
prejudice alone.

\acknowledgments

SH is supported by the Academy of Finland grant 131454 and thanks K. Kohri for 
useful discussions relating to the production and effects of primordial black holes.

\end{document}